\documentclass{mn2e}
 \usepackage[dvips]{epsfig}
 \usepackage{amsfonts,amsbsy}
\title 
  {Decrease in the orbital period of dwarf nova OY Carinae}
\author[J.G.Greenhill {\it et al.}]
{J. G. Greenhill , K. M. Hill, S. Dieters,  K.Fienberg, M.Howlett,
A.Meijers\and  A.Munro, C.Senkbeil\\
       School of Mathematics and Physics, University of Tasmania, 
       Private bag 37, GPO Hobart, Tasmania 7001, Australia} 
      
\date{Draft: 10 July 2006}

\pagerange{\pageref{firstpage}--\pageref{lastpage}}
\pubyear{2006}

\begin{document}

\maketitle

\label{firstpage}

\begin{abstract}
We have measured the orbital light curve of dwarf nova OY Carinae on 8
separate nights between 1997 September and 2005 December. The measurements 
were made in white light using CCD photometers on the Mt Canopus 1 m 
telescope. The time of eclipse in 2005 December was $168 \pm 5$ s earlier than that 
predicted by the Wood et al.(1989) ephemeris. Using the times of eclipse from our 
measurements and the compilation of published measurements by Pratt et al. (1999) 
we find that the observational data are inconsistent with a constant period 
and indicate that the orbital period is decreasing by  $5\pm 1\times 10^{-12}$s/s.
This is too fast to be explained by gravitational radiation emission alone. It is possible 
that the change is cyclic with a period $\sim 35$ years and fractional period change 
$\Delta P/P = 2.6\times 10^{-7}$. This is probably due to solar-cycle magnetic 
activity in the secondary. There are also large systematic deviations, with a time-scale of years,
from a sinusoidal modulation.

\end{abstract}
\begin{keywords}
binaries: close -- stars: evolution -- stars:dwarf novae -- stars:individual: OY Car -- 
gravitational waves -- stars:magnetic fields 
\end{keywords}

\section{Introduction}
OY Car is an eclipsing dwarf nova of the SU UMa class. It contains a 
Roche lobe filling secondary transferring matter via an accretion disc 
onto a white dwarf. The matter flow onto the disc is via a well defined 
stream giving rise to a hot spot on the disc. The orbital period is 
$\sim 91$ minutes (Vogt et al. 1981); below the period gap for cataclysmic 
variables. The light curve shows evidence of 
eclipses of both the white dwarf and of the hot spot by the secondary.

It is generally believed that cataclysmic variables below the period 
gap can lose angular momentum only by emission of gravitational 
radiation (see e.g. Ritter and Kolb 1992). This is a slow process with 
characteristic time-scale $P/P_{dot} \sim 10^{10}$ yr. Many such systems,
however, have long term mass transfer rates (and hence angular momentum 
loss rates) as much as 6 times the rate predicted by gravitational 
radiation emission suggesting that magnetic braking can occur even in 
systems with fully convective secondaries (Warner 1995). 

Many CVs exhibit quasi-periodic variations in their
orbital period and/or optical lightcurves. This cyclic behaviour is thought to be 
caused by solar-cycle-type magnetic activity in the secondary star 
(Warner et al. 1988; Applegate 1992; Richman, Applegate \& Paterson  1994). 
Typical magnetic cycle periods are in the range 4 to 30 years (Baptista et al. 2003). There 
is increasing evidence that this activity occurs even in CVs below the priod gap with fully 
convective companion stars (Ak, Ozkan \& Mattei (2001) and references therein). 
For example the SU UMa star, Z Cha, has a $\sim 107$ minute orbital period which 
shows fractional changes, $\delta P/P \sim 4.4 \times 10^{-7}$ in a 28 yr 
cycle (Baptista et al. 2002). The fractional cyclic changes in the shorter orbital period CVs are, 
however, systematically smaller than those in the longer period systems above the period gap
(Baptista et al 2003). Strictly periodic changes in the orbital period can also be caused by a 
third body in the system causing the centre of mass of the CV to move in and out 
of the plane of the sky but we are not aware of any published evidence for such systems. 

Several ephemerides for OY Car have been published (Vogt et al. 1981; Cook
1985; Wood et al. 1989). The ephemeris by Cook (1985) included a second 
order term suggesting that the orbital period was decreasing.
A later ephemeris (Wood et al. 1989) and a recent study by Pratt et al. (1999) 
using data gathered over 19 years show, however, that there was, at that time, 
no evidence for a decreasing period. 

In this paper we describe eclipse timing measurements of OY Car made 
between 1997 September and 2005 December. We use these data in 
conjunction with the data measured and collated by Pratt et al. (1999)
to show that the ephemeris is now inconsistent with a linear function 
and that the orbital period is decreasing.

\section{Observations}
The observations described in this paper were made using the  
1-m telescope at the University of Tasmania Mt. Canopus Observatory. An 
SBIG ST6 CCD camera was used for the data taken in 1997. All other 
observations were made using an SITe 512 x 512 pixel, thinned, backside 
illuminated CCD with Leach controller and CICADA operating software. A 
journal of observations 
is given in Table 1. Exposure times were 20 s for the ST6 camera and 10 
or 15 s  with the SITe 
camera. The chip readout time was typically 3 s and the cameras were set
in repeated exposure mode so that continuous coverage, with sampling  
interval 13 to 23 s, was obtained for at least one orbital cycle on each 
night. A clear filter was used in all cases.

The image reduction and analysis was carried out using MIDAS and the DoPHOT  
profile fitting photometry system. The images were dark subtracted, flat-fielded and 
trimmed before calling a Midas control language procedure which carried 
out DoPHOT photometry on each image and generated a table of magnitudes 
of OY Car relative to its bright neighbour $\sim 0.4$ arc-minutes to the 
NW. Details of the eclipse portion of a typical light curve are shown in 
Fig 1. 

\begin{figure}
 \epsfig{file=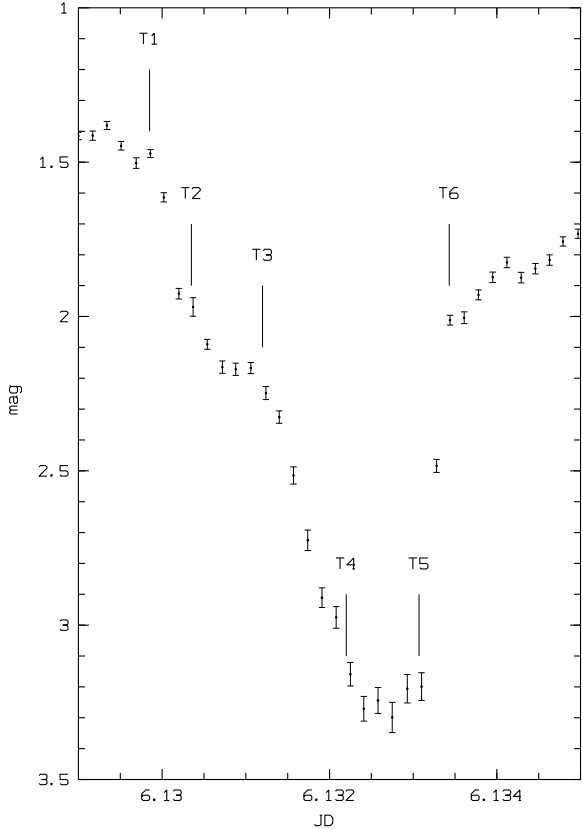, width=8.4cm }
 \caption{Eclipse portion of a typical light curve (2005 December 11). The 
  contact times, T1 - T6, are defined in the text. The times on the
  horizontal axis correspond to JD-2,453,716 at the beginning of each 
  exposure.}
\end{figure}

\section{Results}

\subsection{Eclipse timing}
Eight moments of contact are present in all eclipses in the quiescent 
state. We follow the definitions of Bailey (1979) where T$_1$ and 
T$_2$ correspond to the beginning and end of the primary ingress, 
T$_5$ and T$_6$ are the corresponding times of egress and T$_3$ and 
T$_4$ are the times of hot spot ingress. The eclipse centre 
of the white dwarf primary is given by 
$$ T_0 = 0.25(T_1+T_2+T_5+T_6) $$
To the measured time we must add half the exposure times used on each
night. In order to compare with published ephemerides, the measured times 
were converted to heliocentric times (HJD) and then to Terrestrial 
Dynamic Time. For each eclipse light curve, the resulting Heliocentric 
Julian Ephemeris Date (HJED) is listed in Table 1.

\begin{table}
 \caption{Journal of observations and eclipse timing (O-C in seconds) 
 relative to the ephemeris of Wood et al. 1989.}
 \label{symbols}
 \begin{tabular}{lccc} \hline
  Date  		&  HJED 		& Cycle No.	& O-C\\ \hline
  28.08.97     	& 2450689.04220	& 106074 	& -49$\pm11$\\	              
  28.09.97      	& 2450720.16079 	& 106567	& -47$\pm14$ \\
  06.09.01        & 2452159.00124	& 129362 	& -135$\pm26$ \\       
  02.10.03	& 2452912.98043	& 141307	& -156$\pm10$ \\
  23.12.03    	& 2452997.05749	& 142639 	& -156$\pm10$ \\
  22.05.04 	& 2453147.91661	& 145029	& -147$\pm9$ \\ 
  17.01.05        & 2453388.09108	& 148834	& -163.6$\pm5$ \\
  17.01.05        & 2453388.15424	& 145835	& -159.7$\pm5$ \\ 
  11.12.05        & 2453716.131680	& 154031	& -168.5$\pm5$ \\
  11.12.05        & 2453716.194089	& 154032	& -164.2$\pm5$ \\
  \hline
 \end{tabular}
 \medskip
\end{table}

\subsection{Comparison with published ephemerides.}
\begin{figure}
 \epsfig{file=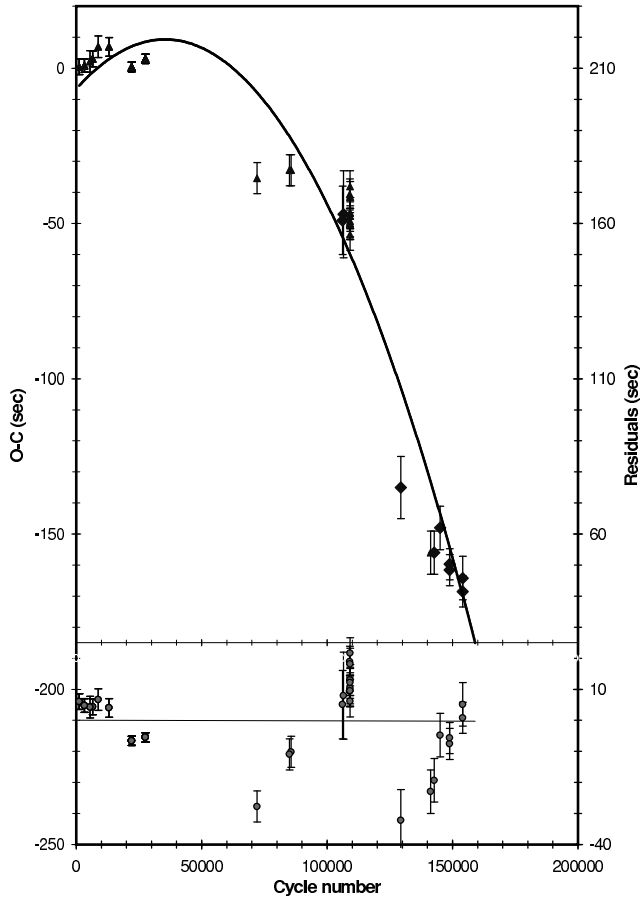, width=8.4 cm }
 \caption{Differences (O-C) between observed and predicted eclipse times 
 (based on the ephemeris of  Wood et al. 1989) vs orbital cycle number.
 Our measurement are represented by the $\diamond$  
 symbols. The solid curve represents the best fitting second order 
 polynomial function (upper panel). The residuals are shown in the lower panel.
 }
\end{figure}

\begin{figure}
 \epsfig{file=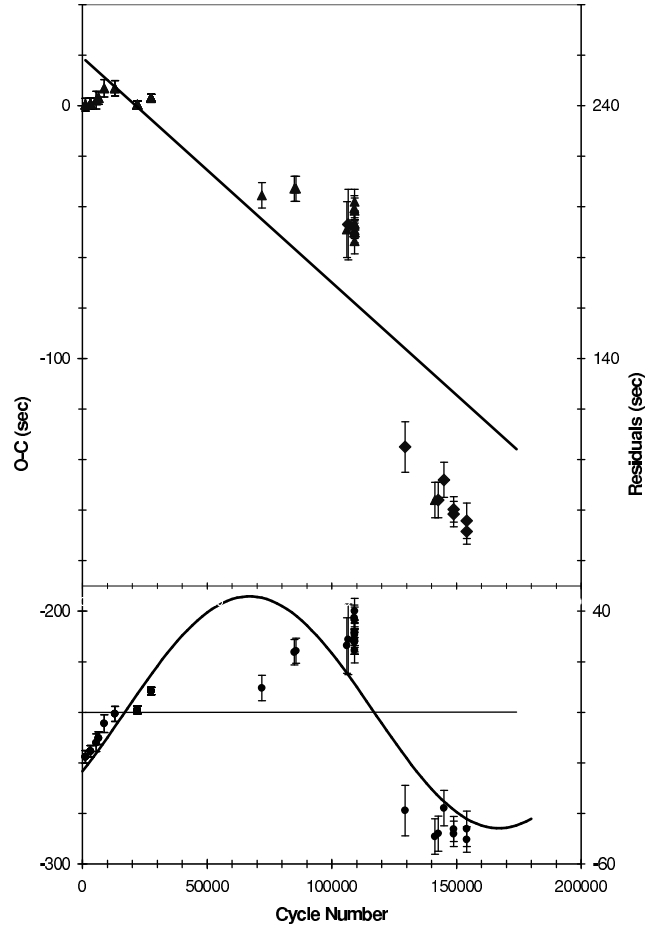, width=8.4 cm }
 \caption{Differences (O-C) between observed and predicted eclipse times 
 (based on the ephemeris of  Wood et al. 1989) vs orbital cycle number .
 Our measurement are represented by the $\diamond$  
 symbols. The solid line (upper panel) represents the best fitting linear function. 
 Residuals to the linear fit and the best fitting sine function are illustrated in the lower panel.
 }
\end{figure}

We compare our eclipse times with predictions from the ephemeris 
of Wood et al (1989) 
$$HJED=(2,443,993.553839\pm 9) + (0.0631209239\pm 5)E $$
where E is the cycle number. The values of E and the differences (O-C) 
between observed and predicted eclipse times  are listed in Table 1.

We combine our results with the compilation of previous measurements in 
Pratt et al. (1999). The variation of O-C with cycle number E since 1979
is illustrated in Fig 2. For the sake of simplicity we have plotted only 
single points representing the means (and error in the mean) where data from this compilation 
were closely spaced in time. In this analysis we have assumed equal precision for each 
measurement with weighting proportional to the number of measurements contributing to each 
mean. 

The data are clearly inconsistent with the assumption of a constant period.  The solid line 
represents the best fitting second order polynomial for all the measurements. 
$$ O-C=-6.6\pm2.6+(9.0\pm1.4 \times 10^{-4})E-(1.3\pm0.1 \times 10^{-8})E^{2} $$
The quadratic term is significant 
at the $12\sigma$ level indicating a period evolution time-scale of $3.7 
\times 10^7$. Cook (1985) found evidence at the 
$3.5\sigma$ level for a period evolution time-scale of $2.1\times 10^7$yr but later 
measurements (Wood et al. 1989) were not consistent with his predictions. We note 
however that, with the passage of time, the quadratic term is becoming increasingly 
dominant so that the constant period hypothesis can be excluded with a high level 
of confidence. 

The fit is poor with $\chi^2/dof=9.4$ for 63 degrees of freedom. There are also highly significant
systematic deviations with time-scales of years. We have  
investigated the possibility that changes in the shape and duration of the eclipse light curve 
might cause these systematic deviations. The mean duration (centre of ingress to egress) for our measurements
is $276.9 \pm 6.0$ s - not significantly different from the value $274.7 \pm 3.4$ s reported 
by Vogt et al. (1981). Similar durations are also evident in the light curves published by Cook (1985), 
Wood et al. (1989) and Pratt et al. (1999). Neither is there any significant variation in the duration of the 
ingress and egress. The phase and duration of hotspot eclipses is much more variable but this cannot 
affect the white dwarf eclipse timing. Hence we conclude that changes in the eclipse light curve do not 
contribute to the observed changes in orbital period. 

Next we investigate the possibility that the period is constant with a sinusoidal modulation.  
In Fig 3 we show a linear fit and residuals. Clearly the linear fit alone is unsatisfactory. We have 
phase folded the data at the 6.3 year period reported for the quiescent magnitude of OY Car 
reported by Ak, Ozkan \& Mattei (2001) but find no correlation. We then used an iterative 
process to determine the best 
sinusoidal fit to the residuals after fixing phase zero at cycle number = 17,000 
($JD \sim 2,445,067$). The best fitting period is $2.0\pm 0.2 \times 10^5$ cycles ($35 \pm 3.5$) 
years with amplitude $46 \pm 3$ s. The solid line in the lower panel of Fig 2 represents 
this periodic modulation. The fit is again poor with $\chi^2/dof=8.1$ for 61 degrees of freedom. 
Application of the F-test indicates however, that the linear plus sinusoid model 
represented in Fig 3 is a better fit than the polynomial model. This is significant at the 99.9 \%
level but, as with the quadratic model, there are highly significant
systematic deviations with time-scales of years.

\section{Discussion}
We consider first the possibility that the decreasing period is due to loss of angular 
momentum from the system and is not due to some cyclic change. Using the 
quadrupole general relativity formula for the fractional rate of change in the orbital 
angular momentum $\dot{J}/J$ (Livio 1994) and the system 
parameters for OY Car (Wood et al. 1989) we find $\dot{J}/J \approx 2.2\times10^{-10}$
per year. The observed rate of change is $\sim 2.7\times 10^{-8}$; almost two orders of 
magnitude faster than the rate predicted for loss by gravitational radiation. 
%Moreover it is an order of magnitude faster than the rate, inferred for dwarf novae with 
%similar orbital periods, from disc luminosity measurements (Warner 1995)   

As already noted the linear plus sinusoid model represented 
in Fig 3 is a better fit than the quadratic although large non-random residuals remain. 
The modulation could be due to the presence in the system of a third object - 
an $M \sim 0.007M_{\odot}$ brown dwarf or massive planet with orbital radius $a \sim9.7$ AU but 
this is very unlikely since most well observed CVs show cyclical period changes (Baptista et al. 2003
and references therein). It is highly improbable that most CVs are members of triple systems.

The most likely cause is quasi-periodic solar-cycle type magnetic activity in the companion. Magnetic 
cycles are inherently less regular and this might be the cause of the departures from the model. The 
fractional period change $\Delta  P/P = 2\pi\Delta(O-C)/P_{mod}$ where $P$ is the orbital period, $P_{mod}$
is the modulation period and $\Delta(O-C)$ is the amplitude of the sinusoidal modulation (Applegate 1992). 
For OY Car we find  $\Delta P/P = 2.6 \times 10^{-7}$, similar to that for other 
short period CVs (Baptista et al. 2003). Departures from the sinusoidal model are however much larger 
than those seen in the cyclic modulations of quiescent magnitude for dwarf novae reported by 
Ak, Ozkan \& Mattei (2001). Baptista et al. (2002) observed similar changes in the timing measurements 
for the SU UMa star Z Cha. 

Finally, we present the ephemerides for the quadratic and sinusoidal modulation models.
For the quadratic model:
\begin{eqnarray*}
HJED & = & 2,443,993.553813+0.0631209343 E\\
& & \mbox{} -(1.47 \times 10^{-13}) E^2
\end{eqnarray*}
For the sinusoidal model:
\begin{eqnarray*}
HJED & = & 2,443,993.55406+0.0631209126 E \nonumber\\
& & \mbox{} + (5.3 \times 10^{-4})sin\frac {2\pi(E-1.7\times 10^{-4})}{2\times 10^5} \nonumber
\end{eqnarray*}
Timing measurements during the next few years should determine which model is valid.

\section{Conclusions}
The orbital period of OY Car is decreasing at an average rate of  $5 \pm 1\times  10^{-12}$  
s/s. This is $\sim 2$ orders of magnitude too fast to be caused by gravitational radiation emission 
alone and loss of angular momentum via a stellar wind is considered unlikely. 

It is likely that the apparent change is cyclic with a period $P = 35 \pm 3.5$ years and 
fractional period change $\Delta P/P = 2.6 \times10^{-7}$. This is probably due to solar-cycle type magnetic 
activity in the secondary. Large irregular deviations 
from the general trend, with time-scales of years, also occur. Measurements over the next few years 
will clarify whether the change in period is cyclic or a continuing decrease.  We find no evidence for 
the 6.3 year modulation in quiescent magnitude reported by Ak, Ozkan \& Mattei (2001).

\section{Acknowledgements}
We are indebted to Bob Watson for suggesting the project as a research training exercise for third 
year Physics students and to the referee Pierre Maxted for many helpful suggestions. We gratefully 
acknowledge financial support for the Mt Canopus Observatory by Mr David Warren.

\bsp % ``This paper has been produced using the ...''

\label{lastpage}

\end{document}